\documentclass[twocolumn,showpacs]{revtex4}
\usepackage{eurosym}
\usepackage{graphicx}
\usepackage{dcolumn}
\usepackage{bm}
\usepackage{amsmath}
\usepackage{amsfonts}
\usepackage{amssymb}

\providecommand{\U}[1]{\protect\rule{.1in}{.1in}}

\begin{document}

\title{Fractional dynamics in the L\'evy quantum kicked rotor}
\author{A. Romanelli}
\altaffiliation[Corresponding author.\\
]{\textit{E-mail address:} alejo@fing.edu.uy}
\affiliation{Instituto de F\'{\i}sica, Facultad de Ingenier\'{\i}a\\
Universidad de la Rep\'ublica\\
C.C. 30, C.P. 11000, Montevideo, Uruguay}
\date{\today }

\begin{abstract}
We investigate the quantum kicked rotor in resonance subjected to
momentum measurements with a L\'evy waiting time distribution. We
find that the system has a sub-ballistic behavior. We obtain an
analytical expression for the exponent of the power law of the
variance as a function of the characteristic parameter of the L\'evy
distribution and connect this anomalous diffusion with a fractional
dynamics.
\end{abstract}

\pacs{: 03.67.-a, 05.30.Pr, 05.40.Fb, 05.45.Mt} \maketitle

\section{Introduction}

During the last decades it has been possible to obtain samples of
atoms at temperatures in the nano-kelvin range \cite{Cohen} (optical
molasses) using resonant or quasi-resonant exchanges of momentum and
energy between atoms and laser light. This spectacular experimental
progress has been accompanied with the development of the
inter-disciplinary fields of quantum computation and quantum
information. In this frame, the study of simple quantum systems,
such as the quantum kicked rotor (QKR) \cite{Izrailev} and the
quantum walk (QW) \cite{Kempe} are extremely useful as models to
design future codes and computers. The behavior of the QKR has two
characteristic modalities: dynamical localization (DL) and ballistic
spreading of the variance in resonance. These different behaviors
depend on whether the period of the kick is a rational or irrational
multiple of $2\pi $. For rational multiples the behavior of the
system is resonant and the average energy grows ballistically and
for irrational multiples the average energy of the system grows, for
a short time, in a diffusive manner and afterwards DL appears.
Quantum resonance is a constructive interference phenomena and DL is
a destructive one. The DL and the ballistic behavior have already
been observed experimentally \cite{Moore,Kanem}.

In ref.\cite{alejo2} we investigated the QKR in resonant regime and
the usual QW when both systems were subjected to decoherence with a
L\'evy waiting time distribution. In the case of the QKR the model
had two strength parameters whose action alternated in a such way
that the time interval between them followed a power law
distribution. In the case of QW the model used two evolution
operators whose alternation followed the same power law
distribution. We showed that this noise in the secondary resonances
of the QKR and in the usual QW produced a change from ballistic to
sub-ballistic behavior. This change of behavior is similar to that
obtained for both systems when they are subjected to an aperiodic
Fibonacci excitation \cite{alejo1,Ribeiro}. In all the above cases
the sub-ballistic behavior is characterized by the time dependence
of the variance, \emph{i.e.} $\sigma^2 (t)\sim t^{2c}$, with
$0.5<c<1$. In a more recent paper \cite{alejo3} we have studied the
QW subjected to measurements with a L\'evy waiting time distribution
and we found that the system had a sub-ballistic behavior. We also
obtained an analytical expression for the exponent of the power law
of the variance as a function of the characteristic parameter of the
L\'evy distribution.

In this paper we present a simple model that allows an analytical
treatment to understand the sub-ballistic behavior previously
reported in ref. \cite{alejo2}. We shall show that the temporal
sequence of the decoherence, and not its intensity, is the main
cause of this unexpected dynamics. With this aim we investigate the
QKR when measurements are performed on the system with waiting times
between them following a L\'{e}vy power-law distribution. We show
that this type of noise indeed produces sub-ballistic behavior. We
obtain analytically a relation between the exponent of the variance
and the characteristic parameter of L\'{e}vy distribution. These
results are identical to the ones obtained in ref. \cite{alejo3},
showing again another aspect of the similarity between QKR and QW,
as pointed out in previous papers
\cite{alejo1,alejo2,alejo4,alejo5}. In addition the toy model
developed in this work shows that a quantum system in combination
with a L\'{e}vy stochastic process may produce a fractional dynamics
for the averaged behavior.

\section{L\'evy quantum kicked rotor}

The QKR is one of the most simple and best investigated model whose
classical counterpart displays chaos. It has the following Hamiltonian
\begin{equation}
H=\frac{P^{2}}{2I}+K\cos \theta \sum_{n=1}^{\infty }\delta (t-n),
\label{qkr_ham}
\end{equation}%
where $P$ is the angular momentum operator, $I$ is the moment of
inertia, $K$ is the strength parameter, $\theta $ is the angular
position. The external kicks occur at times $t=n$ with $n$ integer
and unity period. In the angular momentum representation, $P|\ell
\rangle =\ell \hbar |\ell \rangle $, the wave-vector is $|\Psi
(t)\rangle =\sum_{\ell =-\infty }^{\infty }a_{\ell }(t)|\ell \rangle
$ and the average energy is $E(t)=\left\langle \Psi \right\vert
H\left\vert \Psi \right\rangle =\varepsilon \sum_{\ell =-\infty
}^{\infty }\ell ^{2}\left\vert a_{\ell }(t)\right\vert ^{2}$, where
$\varepsilon =\hbar ^{2}/2I$. Using the Schr\"{o}dinger equation the
quantum map is readily obtained from the Hamiltonian (\ref{qkr_ham})
\begin{equation}
a_{\ell }({t+1})=\sum_{j=-\infty }^{\infty }U_{\ell j}a_{j}({t}),
\label{mapa}
\end{equation}%
where the matrix element of the time step evolution operator $U(\kappa )$ is
\begin{equation}
U_{\ell j}=i^{-(j-\ell )}e^{-ij^{2}\varepsilon /\hbar }\,J_{j-\ell }(\kappa
),  \label{evolu}
\end{equation}%
$J_{m}$ is the $m$th order cylindrical Bessel function and its argument is
the dimensionless kick strength $\kappa \equiv K/\hbar $. The resonance
condition does not depend on $\kappa $ and takes place when the frequency of
the driving force is commensurable with the frequencies of the free rotor.
Inspection of Eq.(\ref{evolu}) shows that the resonant values of the scale
parameter $\tau \equiv \varepsilon /\hbar $ are the set of the rational
multiples of $2\pi $, \textit{i.e.} $\tau =2\pi $ $p/q$. When $p/q$ is an
integer the resonance is called principal and when it is a non integer
rational it is called secondary.

\begin{figure}[h]
\begin{center}
\includegraphics[scale=0.4]{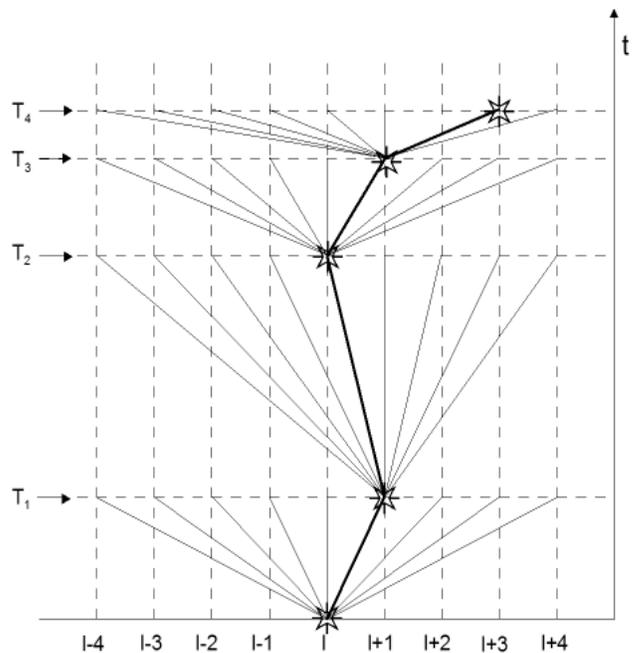}
\end{center}
\caption{Paths of the LQKR wave function as a function of the
angular momentum, $l$. The measurements are performed at times
${T_{i}}$ and the wave function collapses at these times. Between
measurements the system has an unitary quantum evolution}
\label{fig1}
\end{figure}

The dynamics of the L\'{e}vy quantum kicked rotor (LQKR) will be
generated by a large sequence of two time-step unitary operators
$U_{0}$ and $U_{1}$ as was done in a previous work \cite{alejo2}.
But now $U_{0}$ is the ``free" evolution of the QKR in resonance and
$U_{1}$ is the operator that measures the angular momentum of the
QKR. The time interval between two applications of the operator
$U_{1}$ is generated by a waiting-time distribution $\rho (T) $,
where $T$ is a dimensionless integer time step, see Fig.~\ref{fig1}.
The detailed mechanism to obtain the evolution is given in \cite{alejo2}. We
take $\rho (T)$ in accordance with the L\'{e}vy distribution \cite%
{Shlesinger} that includes a parameter $\alpha $, with $%
0<\alpha \leq 2$. When $\alpha <2$ the second moment of $\rho $ is
infinite, when $\alpha =2$ the Fourier transform of $\rho $ is the
Gaussian distribution and the second moment is finite. Then, this
distribution has no characteristic size for the temporal jump,
except in the Gaussian case. The absence of scale makes the L\'{e}vy
random walks scale-invariant fractals. This means that any classical
trajectory has many scales but none in particular dominates the
process. This distribution appears, for example, in quantum optics
\cite{Bardou} as an appropriate tool to describe cooled atomic
samples in terms of a competition between a trapping process (the
atom falls in the optical trap) and a recycling process (the atom
leaves the trap and eventually return to it).
The most important characteristic of the L%
\'{e}vy noise is the power-law shape of the tail, accordingly in this work
we use the waiting-time distribution
\begin{equation}
\rho (t)=\frac{\alpha }{\left( 1+\alpha \right) }\left\{
\begin{array}{cc}
1, & 0\leq t<1, \\
\left( \frac{1}{t}\right) ^{\alpha +1}, & t\geq 1.%
\end{array}%
\right.   \label{Levy}
\end{equation}%
To obtain the time interval $T$ we sort a continuous variable $t$ in
agreement with Eq. (\ref{Levy}) and then we take the integer part
$T_{i}$ of this variable \cite{alejo2}.

In what follows we assume that the resonance condition of the QKR is
satisfied, for the sake of simplicity we take $\varepsilon /\hbar =2\pi $ in
such a way that the operator $U_{0}$ corresponds to the first principal
resonance. This choice does not imply a loss of generality for our results
as we shall show below.

Let us suppose that the wave function is measured at the time $t$,
then it evolves according to the unitary map Eq.~(\ref{mapa}) during
a time interval $T$, and again at this last time $t+T$ a new
measurement is performed. In Fig.~\ref{fig1} we present a path
diagram of the state evolution. It shows four time steps when the
measurements are perform, between measurements there is an unitary
evolution. When the measurement is performed the wave function
collapses in a momentum state. The resulting states after successive
measurements need not be contiguous states as in the QW because all
transitions are possible.

In the figure we present a generic and arbitrary path with bold
line. From this diagram we can write a dynamical equation for the
probabilities of the LQKR momenta. To begin note that the
probability that the wave function collapses in the eigenstate
$|j\rangle $, due to a momentum measurement, starting from the
eigenstate $|0\rangle $ after a time $T$ is
\begin{equation}
P_{j}(T)\equiv {\vert a_{j}({T}) \vert\ }^2. \label{position}
\end{equation}
The momentum distribution $P_{j}$ depends on the initial state and
on the time interval $T$ because of the collapse of the wave
function, and it will play the role of transition probabilities for
the global evolution. The mechanism used to perform momentum
measurements assures that these distributions will repeat themselves
around the new momentum. Then it is straightforward to build the
probability distribution $P_{j}$ at the new time $t+T$ as a
convolution between this distribution at the time $t$ and the
conditional probability:
\begin{equation}
P_{l}(t+T)=\sum\limits_{j=-\infty}^{\infty}q_{l-j}(T)P_{j}(t),
\label{markov1}
\end{equation}
where $q_{l-j}$ are the transition probabilities from state $j$ to
state $l$ and the sum is extended between $-\infty$ and $+\infty$
because all the transition are possible. To calculate $q_{l-j}$ the
original dynamical equations Eqs.~(\ref{mapa}, \ref{evolu}) and the
properties of the Bessel function are used to obtain a connection
between the initial pure state after a measurement and all possible
final states before the next measurement
\begin{equation}
q_{l-j}(T)={(J_{j-l }(\kappa T))}^2.  \label{transition}
\end{equation}
The Eq.~(\ref{markov1}) is a sort of master equation, but not strictly
because of the time dependence of the transition probabilities.

There are many ways of solving Eq.~(\ref{markov1}), we choose the
method of the generating function $G(z.t)$ defined as
\begin{equation}
G(z,t)=\sum\limits_{j=-\infty}^{\infty}z^jP_{j}(t),  \label{genera}
\end{equation}
where we shall take the auxiliary variable $z$ as $z\equiv
e^{i\varphi}$, with $\varphi$ real. It is easy to prove using
Eq.~(\ref{markov1}), that
\begin{equation}
G(z,t+T)=G(z,t)J_{0}(2\kappa T sin \varphi/2),  \label{genera}
\end{equation}
The first and second moments are given by
\begin{align}
m_{1}(t)\equiv G^{\prime }(1,t) ,  \label{mom1a} \\
m_{2}(t)\equiv G^{\prime \prime }(1,t),  \label{mom2a}
\end{align}
where the prime indicates differentiation with respect to $z$. Then
using these equations and Eq.~(\ref{genera}) the following maps for
the moments are obtained:
\begin{align}
m_{1}(t+T)& =m_{1}(t)+m_{1q}(T),  \label{mom1} \\
m_{2}(t+T)& =m_{2}(t)+2m_{1}(t)m_{1q}(T)+m_{2q}(T),  \label{mom2}
\end{align}%
where
\begin{align}
m_{1q}(T)&=\sum\limits_{l=-\infty}^{l=\infty}lq_{l}(T),  \label{mom1} \\
m_{2q}(T)&=\sum\limits_{l=-\infty}^{l=\infty}l^{2}q_{l}(T).
\label{mom2}
\end{align}
Note that $m_{1q}(T)$ and $m_{2q}(T)$ are the first and second moments of
the unitary evolution between measurements. From these expressions and using
the Eq.(\ref{transition}) the following results are obtained $m_{1q}(T)=0$
and $m_{2q}(T)=\frac{1}{2}\kappa T^2$. Therefore the global variance $\sigma
^{2}(t)=m_{2}(t)-m_{1}^{2}(t)$ verifies that
\begin{equation}
\sigma ^{2}\left( t+T\right) =\sigma ^{2}\left( t\right) +\sigma _{q}^{2}(T),
\label{varianza}
\end{equation}
where $\sigma _{q}^{2}(T)=m_{2q}(T)-m_{1q}^{2}(T)=\frac{1}{2}\kappa T^2$ is
the variance associated to the unitary evolution between measurements. Note
that the value of the coefficient of $T^2$ is a consequence of using the
principal resonance but the time dependence remains unchanged for any other
resonance. From these last equations is easy to show that
\begin{equation}
\sigma ^{2}\left( t\right) =\frac{1}{2}\kappa N\sum\limits_{i=1}^N T_{i}^2,
\label{varianza2}
\end{equation}
where
\begin{equation}
t=\sum\limits_{i=1}^N T_{i},  \label{time}
\end{equation}
and $N$ is the number of measurements performed. These results are generic,
now we shall calculate the average of Eq. (\ref{varianza2})
\begin{equation}
\left\langle \sigma ^{2}\left( t\right) \right\rangle = \frac{1}{2}\kappa t
\frac{\left\langle T_{i}^{2}\right\rangle}{\left\langle T_{i}\right\rangle},
\label{varianza3}
\end{equation}
where the relation $t={\left\langle T_{i}\right\rangle}{\left\langle
N\right\rangle}$ was used. The first and the second moments of the waiting
time for our L\'{e}vy distribution, Eq. (\ref{Levy}), are
\begin{equation}
\left\langle T_{i}\right\rangle =\frac{\alpha }{\alpha +1}\left\{
\begin{array}{cc}
\left( \frac{1}{2}+\frac{t^{1-\alpha }-1}{1-\alpha }\right) , & \alpha \neq
1, \\
\left( \frac{1}{2}+\ln (t)\right) , & \alpha =1,%
\end{array}%
\right.  \label{first}
\end{equation}
\begin{equation}
\left\langle T_{i}^{2}\right\rangle =\frac{\alpha }{\alpha +1}\left\{
\begin{array}{cc}
\left( \frac{1}{3}+\frac{t^{2-\alpha }-1}{2-\alpha }\right) , & \alpha \neq
2, \\
\left( \frac{1}{3}+\ln (t)\right) , & \alpha =2.%
\end{array}%
\right.  \label{secon}
\end{equation}%
Substituting these expressions in Eq.~(\ref{varianza3}) and for a large time
\begin{equation}
\left\langle \sigma ^{2}\left( t\right)
\right\rangle=\frac{1}{2}\kappa \left\{
\begin{array}{cc}
t^2\,, & \text{if \ }0\leqslant \alpha \leqslant 1, \\
t^{(3-\alpha )}\,, & \text{if \ }1\leqslant \alpha \leqslant 2.%
\end{array}%
\right.  \label{limite}
\end{equation}%
Therefore when $t\rightarrow \infty $ the variance behaves as
$\left\langle \sigma ^{2}\left( t\right) \right\rangle\sim t^{2c}$
where
\begin{equation}
c=\left\{
\begin{array}{cc}
1\,, & \text{if \ }0\leqslant \alpha \leqslant 1, \\
\frac{1}{2}(3-\alpha )\,, & \text{if \ }1\leqslant \alpha \leqslant 2.%
\end{array}%
\right.  \label{limite2}
\end{equation}%
This result shows that measurements do not break completely the
coherence of the system on a time scale that includes several of
them. For $0\leqslant \alpha \leqslant 1$ the ballistic behavior is
preserved as in the usual resonant QKR, and for $1\leqslant \alpha
\leqslant 2$ it is lost and the sub-ballistic behavior takes place.
When $\alpha=2$ the system has a diffusive behavior as in the usual
Brownian motion. From the fact, that the exponent $c$ does not
depend on $\kappa$ it follows that the results are valid for both
primary and secondary resonances. However the coefficient $c$
depends on the microscopic law of evolution,
$\sigma_{q}^{2}(T)\propto T^2$. Then, we may pose the question if
there exists a relation between the time dependence of
$\sigma_{q}^{2}(T)$ in the quantum unitary evolution between
measurements and the exponent $c$ of the power law for the averaged
variance. To answer this question we shall suppose a unitary quantum
evolution that produces the following variance
\begin{equation}
\sigma _{q}^{2}(T)\propto T^{\beta }, \label{varianza5}
\end{equation}
where $\beta$ is a constant. The reasoning to obtain the exponent
$c$ can now be repeated, it is only necessary to calculate again the
new expression for a general $\beta$ moment with the L\'{e}vy
waiting time distribution, that is
\begin{equation}
\left\langle T_{i}^{\beta}\right\rangle =\frac{\alpha }{\alpha
+1}\left\{
\begin{array}{cc}
\left( \frac{1}{\beta+1}+\frac{t^{\beta-\alpha }-1}{\beta-\alpha
}\right) , & \alpha \neq
\beta, \\
\left( \frac{1}{\beta+1}+\ln (t)\right) . & \alpha =\beta%
\end{array}%
\right.  \label{beta}
\end{equation}%
Then, in this generic case, for $t\rightarrow \infty $, the exponent
$c$ is
\begin{equation}
c=\frac{1}{2}\left\{
\begin{array}{cc}
\beta \,, & \text{if \ }\alpha\leqslant \beta ,0\leqslant \alpha \leqslant 1, \\
(\beta-\alpha+1) \,, & \text{if \ }\alpha\leqslant \beta ,1\leqslant \alpha \leqslant 2, \\
\alpha \,, & \text{if \ }\beta\leqslant \alpha ,0\leqslant \alpha \leqslant 1, \\
1\,. & \text{if \ }\beta\leqslant \alpha ,1\leqslant \alpha \leqslant 2 \\
\end{array}%
\right.  \label{limite3}
\end{equation}%
This expression shows that these systems can exhibit diffusive,
sub-diffusive, ballistic or sub-ballistic behaviors depending on the
values of $\alpha$ and $\beta$. Then, in the theoretical frame of
fractional dynamics \cite{sokolov} Eq.~(\ref{markov1}) together with
the L\'{e}vy distribution would generate a generalized master
equation from which a generic fractional diffusion equation
\cite{metzler} could be built. This fractional dynamics approach has
as an extreme case the classical diffusion equation for
$\alpha=\beta=2$.
\section{Conclusion}
\label{sec:conclusion} 
The quantum resonances of the QKR\ have been experimentally observed
in samples of cold atoms interacting with a far-detuned standing
wave of laser light \cite{Moore,Ammann}. We study the QKR subjected
to measurements with a L\'{e}vy waiting-time distribution. As the
Gaussian distribution is a particular case of the L\'{e}vy
distribution, then our study is open to wider experimental
situations. We showed numerically \cite{alejo2} that a L\'{e}vy
noise does not break completely the coherence in the dynamics of the
QKR but produces a sub-ballistic behavior.  There the system was
also a LQKR but the operators $U_{0}$ and $U_{1}$ corresponded to
the same secondary resonance with two different values of the
strength parameter $\kappa$ and these operators do not commute. It
is important to note that if the operators correspond to a primary
resonance the ballistic behavior was retained due to the
commutativity between the operators $U_{0}$ and $U_{1}$
\cite{alejo2}.  In the present model one of the operators is
unitary, and may correspond to any resonance of the QKR, and the
other is the measurement operator. These operators also do not
commute and again this is linked to the sub-ballistic behavior. Then
we can conclude that for the LQKR the behavior of $\sigma(t)$
depends on the commutativity and the waiting time distribution, both
models show the same physics. We developed a simple analytical
theory to connect the waiting time parameter $\alpha$ with the
exponent $c$. The LQKR behaves like the QW subjected to the same
measurement process \cite{alejo3}, strengthening our previously
established parallelism between both systems
\cite{alejo1,alejo2,alejo4,alejo5}, where the resonant QKR is
interpreted as a QW in momentum space. The type of model developed
in this work shows that a quantum system in combination with a
L\'{e}vy stochastic process leads to an anomalous diffusion and not
to the well known diffusive process of Browniam motion. Finally,
this simple toy model may help to understand the connection between
a fractional approach and a generalized master equation.

\label{sec:Acknowledgments} I thank V. Micenmacher for stimulating
discussions and comments. I acknowledge support from PEDECIBA and
PDT S/C/IF/54/5.

\end{document}